\documentclass[twocolumn]{revtex4-2}

\usepackage[utf8]{inputenc}
\usepackage{amsmath}
\usepackage{amsfonts}
\usepackage{graphicx}
\usepackage{tikz}
\usetikzlibrary{quantikz2}

\begin{document}

    \title{A simpler Gaussian state-preparation}

    \author{Parker Kuklinski}
    \email{Parker.Kuklinski@ll.mit.edu}
    \affiliation{MIT Lincoln Laboratory, Lexington, MA 02139, USA}
    \author{Benjamin Rempfer}
    \email{Benjamin.Rempfer@ll.mit.edu}
    \affiliation{MIT Lincoln Laboratory, Lexington, MA 02139, USA}
    \author{Kevin Obenland}
    \affiliation{MIT Lincoln Laboratory, Lexington, MA 02139, USA}
    \author{Justin Elenewski}
    \affiliation{MIT Lincoln Laboratory, Lexington, MA 02139, USA}
    
    \date{\today}
	
	%============================== ABSTRACT
    \begin{abstract}
The ability to efficiently state-prepare Gaussian distributions is critical to the success of numerous quantum algorithms. The most popular algorithm for this subroutine (Kitaev-Webb) has favorable polynomial resource scaling, however it faces enormous resource overheads making it functionally impractical. In this paper, we present a new, more intuitive method which uses exactly $n-1$ rotations, $(n-1)(n-2)/2$ two-qubit controlled rotations, and $\lfloor (n-1)/2\rfloor$ ancilla to state-prepare an $n$-qubit Gaussian state. We then apply optimizations to the circuit to render it linear in T-depth. This method can be extended to state-preparations of complex functions with polynomial phase. 
\end{abstract}

\keywords{State preparation, Gaussian distribution, circuit optimization}
    %============================== 
		
	\maketitle
    
    %============================== MAIN BODY
    \section{Introduction}
\label{Introduction}

State-preparation \cite{soklakov06, plesch11} is an essential ingredient for many quantum algorithms spanning quantum chemistry \cite{babbush18, fomichev24}, machine learning \cite{sierra_sosa20}, financial mathematics \cite{stamatopoulos24}, linear system solving \cite{clader13}, and differential equations \cite{krovi23}. In general, state-preparation is a difficult task that requires an exponential number of resources \cite{aaronson15} (i.e. a generic $n$-qubit state can have $N=2^n$ unique elements). Many procedures exist for tackling this problem, but necessarily cannot overcome the inherent exponential cost of producing an arbitrary state (e.g. one can produce a state via a circuit with linear depth at the spatial cost of exponentially many qubits \cite{zhang22}).

Often, however, this worst-case scenario is avoided in practice. Many times we wish to produce a state which has a simple functional representation such as $|\psi\rangle =\sum _{x=0}^{N-1}f(x)|x\rangle$. If the function is simple to write down, one may hope that this brevity translates to an efficient state-preparation scheme. One example of this is the phase gradient state \cite{kitaev02} in which a stack of the appropriate single-qubit Y-rotations implements a state proportional to $\sum _{x=0}^{N-1}\alpha ^x|x\rangle$ (see Figure \ref{linear real phase}).

Much of the quantum state-preparation literature attempts to capture large classes of functions with generic procedures. The most cited of these is the Grover-Rudolph algorithm \cite{grover02} which posits a method to load functional states so long as the functions are efficiently integrable. Though this method has persisted for several decades, recently many authors have commented on its impracticality both due to large quantum and classical costs \cite{chakrabarti21, herbert21}. Other generic state-preparation schemes include those for smooth functions \cite{holmes20}, piecewise continuous functions \cite{wang21, guseynov24}, functions with favorable Taylor or Fourier decompositions \cite{rosenkranz24, zylberman24}, quantum logic gate methods \cite{lemieux24}, and machine-learning based variational algorithms \cite{arrazola19}. While these methods are attractive for their wide applicability, in practice there are many overheads which render them infeasible for producing sufficiently accurate states. Especially when applying these broad methods to the state-preparation of elementary functions, it would seem there is efficiency left on the table that could be captured by an algorithm which exploits mathematical properties of the particular function.

One state-preparation function which has arguably received the most attention is the Gaussian distribution due to its ubiquity in physics and mathematics, including its application to factoring \cite{regev25}, lattice cryptography \cite{chen24}, Monte Carlo simulation \cite{chakrabarti21}, and Hamiltonian simulation \cite{ward09}. However, a review of the Gaussian state-preparation literature reveals that current methods have not adequately used the specificity of the Gaussian to gain efficiency over the generic methods already in existence. The most cited procedure for Gaussian state-preparation is the Kitaev-Webb algorithm \cite{kitaev08} which elucidates a recursive procedure derived from Grover-Rudolph. Though the algorithm has polynomial complexity several more recent papers have elaborated on the enormous costs of the algorithm; Deliyannis \textit{et.\ al.} \cite{deliyannis21} show that Kitaev-Webb requires tens of thousands of CNOT gates to state-prepare a one-dimensional Gaussian on ten qubits (costs arising from coherently evaluating Jacobi functions), neglecting to comment on the likely proportionally large rotation gate overhead. Others, including Lemieux \textit{et.\ al.} \cite{lemieux24}, Bagherimehrab \textit{et.\ al.} \cite{bagherimehrab22}, and Rattew \textit{et.\ al.} \cite{rattew21} opt for a `rejection sampling' technique, preparing initial approximation states and using quantum logic gates to construct a comparator operator and recover the Gaussian. Again, these methods are general and correspondingly do not take advantage of specific mathematical structure in the function; \cite{lemieux24} documents costs of state-preparing Gaussian distributions on the order of hundreds of thousands of T-gates and a large number of ancilla to enact the comparators. Still further, other papers abandon the notion of a structured approach altogether and use machine learning to optimize small variational circuits preparing approximations of Gaussian states \cite{iaconis24, manabe24}. While these methods produce states with poor accuracy and have no scaling guarantees, the circuits are reasonably compact and can be implemented on NISQ hardware.

It is in this context that we present a novel state-preparation procedure for the Gaussian distribution, one which exploits particular mathematical structure of the Gaussian function. To understand the algorithm, notice every integer has a binary decomposition in a logarithmic number of bits. Further, every square integer can be represented as a summation of a polylogarithmic number of pairs of binary inputs. Thus an exponential of a square integer $\alpha ^{x^2}$ (i.e. evaluating a normal distribution at an input $x$) can be represented as a \emph{product} of exponentials of these pairs of binary inputs. This form has a direct analogy to quantum circuits, where the constituent exponentials of pairs of binary inputs can be expressed as doubly controlled (controls on the qubits of the binary inputs) rotation gates targeted on an ancilla register. Building out this circuit leads to a block-encoding of a Gaussian distribution on the diagonal of a matrix (i.e. a Gaussian window operator) which we can apply to a uniform superposition (or any other state of interest) and measure off the ancilla qubits to recover the Gaussian state. We then enact several modifications to optimize the circuit for expected T-depth, needing only approximately $2,000$ T-depth to prepare a $20$ qubit Gaussian state to $L^2$ error $10^{-9}$; this is approximately a $100\times$ improvement over similar figures in the literature.

The benefits of this approach beyond the favorable T-depth are numerous, the most intriguing of which is the simplicity of the circuit. To prepare an $n$-qubit Gaussian one only needs $n-1$ single-qubit rotation gates, $(n-1)(n-2)/2$ doubly controlled rotation gates, and $\lfloor (n-1)/2\rfloor$ ancilla qubits, moreover the rotation angles are extremely simple to compute compared to existing methods. Further, the circuits are so simple that we expect it may be possible to run them on NISQ hardware with reasonable accuracy and offer improvements over current ML-based approaches. These circuits also offer an immediate extension to state-preparation of multi-dimensional Gaussian states, as well as states with complex polynomial phases $\sum _{x=0}^{N-1}e^{i\alpha x^d}|x\rangle$ and stretched exponential states $\sum _{x=0}^{N-1}\alpha ^{x^d}|x\rangle$. Furthermore, complex polynomial phase circuits can be combined with an LCU to construct state-preparations of sinusoids of polynomial phase with constant T-depth.

The remainder of this paper is structured as follows; Section \ref{Definitions} introduces relevant definitions and concepts related to state-preparation and discrete Gaussian distributions. Section \ref{State-preparation circuits} outlines the basic circuit construction, Section \ref{Circuit Optimizations} discusses certain optimizations that can be made, and Section \ref{Resource Estimation} documents explicit resource estimates.
    \section{Definitions}
\label{Definitions}

We begin by reviewing definitions essential to the arguments below, first a consistent definition of state-preparation.

\subsection{State-Preparation}
\label{State-preparation}

\begin{figure}
\input{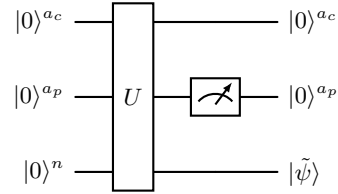}
\caption{A state-preparation circuit $U$ for $|\psi\rangle$ using $a_c$ clean ancilla and $a_p$ persistent ancilla, resulting in an approximation state $|\tilde{\psi}\rangle$.}
\label{cpfig}
\end{figure}

In state-preparation we construct a circuit $U$ to produce an approximation to a desired $n$-qubit quantum state $|\psi\rangle$, call this approximation $|\tilde{\psi}\rangle$. The circuit $U$ depicted in Figure \ref{cpfig} may require $a_c$ \emph{clean ancilla} which begin and end in the $|0\rangle$ as well as $a_p$ \emph{persistent ancilla} which must be measured in the $|0\rangle$ state to successfully recover $|\tilde{\psi}\rangle$ in the data register. In other words, we have
\begin{equation}
\label{state-preparation}
    U|0\rangle ^{\otimes (a_c+a_p+n)}=|0\rangle ^{\otimes a_c}\left(\gamma |0\rangle ^{\otimes a_p}|\tilde{\psi}\rangle +\sqrt{1-|\gamma|^2}|g\rangle\right)    
\end{equation}
where $\gamma$ is the \emph{subnormalization factor} and $|g\rangle$ is an $(a_p+n)$ -- qubit garbage state. If we na\"{i}vely measure the persistent ancilla, the probability of a successful measurement is given by $|\gamma |^2$ and thus we would expect to repeat the entire operation $1/|\gamma |^2$ times to recover $|\tilde{\psi}\rangle$. This could be quadratically mitigated by using amplitude amplification, however the subnormalization factors we will encounter are too large to benefit from the additional overhead introduced by amplitude amplification circuits. We define the error of the approximation by computing the $L^2$ norm $\epsilon =\lVert |\psi\rangle -|\tilde{\psi}\rangle\rVert$. The quantity we are ultimately interested in optimizing is the \emph{expected T-depth} \cite{niemann19} to produce a state with accuracy $\epsilon$; the notion of expectation must be included since state-preparation circuits with non-unit subnormalization have a chance of failing and may need to be repeated. If $U$ has a T-depth of $T$ and a subnormalization factor of $\gamma$, then repeating-until-success (RUS) yields an expected T-depth of $T/|\gamma |^2$.

\subsection{T-gate Sources}
\label{T-gate sources}

\begin{figure}
\input{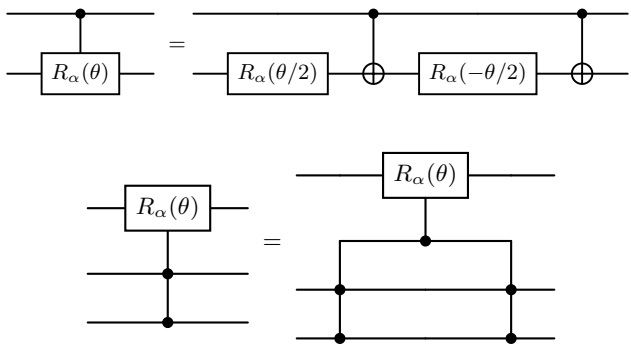}
\caption{Circuit decompositions for controlled-rotations with respect to an arbitrary axis $\alpha$ (\emph{Top}) Decomposition of single-qubit controlled rotation into two uncontrolled rotations (\emph{Bottom}) Decomposition of two-qubit controlled rotation into a Toffoli pair and a single-qubit controlled rotation.}
\label{controlled rotation figure}
\end{figure}

In the circuits below, T-gates will only arise from Toffoli gates and rotation gates. From \cite{bocharov15}, any single-qubit rotation gate regardless of angle can be synthesized to $L^2$ error $\epsilon$ using approximately $1.15\log _2(1/\epsilon )+9.2$ T-gates. The circuit in Figure \ref{controlled rotation figure} shows a decomposition of a controlled rotation into two single-qubit rotations; by approximating each of these to $\epsilon /2$ accuracy the controlled rotation is approximated to $\epsilon$ accuracy and costs $2(1.15\log _2(2/\epsilon )+9.2)=2.3\log _2(1/\epsilon )+20.7$ T-gates. For a doubly controlled rotation, we use the construction in Figure \ref{controlled rotation figure} decomposing into two Toffoli gates and a controlled rotation. From \cite{jones13}, a Toffoli pair costs 4 T-gates to implement, thus the total cost is $2.3\log _2(1/\epsilon )+24.7$. However, as we will be primarily concerned with T-depth, we can execute the 4 T-gates of the Toffoli pair simultaneously with the first 4 T-gates from the controlled rotation, so the T-depth of a doubly controlled rotation is the same as that of a single-qubit controlled rotation.

For convenience we will list several commonly occurring rotation gates. For some value $|\alpha |<1$ we define:

\begin{widetext}
\begin{equation}
\label{A_m}
    A_m=R_y\left(2\cos ^{-1}\left(\frac{1}{\sqrt{1+\alpha ^{2^{m+1}}}}\right)\right) =\frac{1}{\sqrt{1+\alpha ^{2^{m+1}}}}\begin{pmatrix} 1 & -\alpha ^{2^m} \\ \alpha ^{2^m} & 1\end{pmatrix},
\end{equation}
\begin{equation}
\label{B_m}
    B_m=R_y\left(2\cos ^{-1}\left(\alpha ^{2^m}\right)\right) =\begin{pmatrix} \alpha ^{2^m} & -\sqrt{1-\alpha ^{2^{m+1}}} \\ \sqrt{1-\alpha ^{2^{m+1}}} & \alpha ^{2^m}\end{pmatrix}.
\end{equation}
\end{widetext}

We use $A_m$ (occasionally written as $A(m)$ for non-integer $m$) to induce a ratio of $\alpha ^{2^m}$ between the bottom-left entry and the top-left entry, and $B_m$ to encode $\alpha ^{2^m}$ in the top-left entry. If $\alpha$ is sufficiently close to 1 and $m$ is small, $A_m$ approaches $XH$ and $B_m$ approaches identity. Conversely, as $m$ increases $A_m$ approaches identity and $B(m)$ approaches an $X$ gate. For a related state-preparation procedure for functions of polynomial phase, we will make use of the $Z$ rotation
\begin{equation}
\label{Z_m}
    Z_m=R_z(\alpha 2^m)=\begin{pmatrix} 1 & 0 \\ 0 & e^{i\alpha 2^m}\end{pmatrix}
\end{equation}
which encodes a diagonal difference with a $e^{i\alpha 2^m}$ phase difference between top-left and bottom-right entries. Notice that $Z_m$ is a standard rotation when uncontrolled and can be implemented via a Fredkin pair and standard rotation in the controlled setting \cite{fredkinCZ}.

\subsection{Discrete Gaussian Distribution}
\label{Discrete Gaussian distribution}

The primary focus of our paper is producing an $n$-qubit quantum state with amplitudes proportional to a Gaussian distribution. For convenience, we will omit reference to Euler's constant and instead use $\alpha$ ($\alpha <1$) as the exponential base. For an infinite Gaussian distribution, one may be inclined to write the target state as $|\psi\rangle\propto\sum _{x=-\infty}^{\infty}\alpha ^{x^2}|x\rangle$ where the basis state index $x$ ranges over all integers and is thus centered at $x=0$. However, since we will always have access to an even number of basis states, we will offset the input by $1/2$ and instead target a `half-Gaussian' state $|\psi\rangle\propto\sum _{x=0}^{\infty}\alpha ^{(x+1/2)^2}|x\rangle$. To symmetrize this into a full Gaussian we apply a Hadamard gate on an additional qubit (creating two copies of the half Gaussian, one facing the wrong direction) and a series of CNOT gates with open control on the top qubit to flip the correct half of the state.  That these target states are infinite do not offer an existential issue since they can be approximated by a state over finitely many qubits.

Reporting results on these state-preparation schemes offers an inherent challenge because there are many ways these distributions could transform as the number of qubits increases. If we keep $\alpha$ constant, then increasing the number of qubits (i.e. increasing the size of the domain) will be effectively inconsequential past a certain point since the state will contain amplitude extremely close to zero; given an accuracy threshold there is no reason to interact with an increasing number of these qubits. Alternatively one may wish the granularity of the discrete sampling to increase with an increasing number of qubits. In some cases, one may wish to produce a Gaussian state over a fixed window, i.e. targeting an $n$-qubit state $|\psi\rangle\propto\sum _{x=0}^{N-1}\beta ^{(x/(N-1)-1/2)^2}|x\rangle$ where $N=2^n$. Here, $1/\sqrt{\beta}$ is the ratio of the largest amplitude to the smallest amplitude; ideally we would like this quantity to be large to capture a significant proportion of the Gaussian distribution (e.x. $\beta\approx 1.4\times 10^{-11}$ captures 5 standard deviations). We will clearly state which quantum state is being targeted in the results.
    \section{State-preparation Circuits}
\label{State-preparation circuits}

We now present state-preparation circuits for a class of functions including complex functions with linear and polynomial phase as well as exponential and Gaussian functions; these can easily be extended to complex functions of polynomial phase and stretched exponential functions \cite{cardona07}. We defer explicit resource analysis to Section \ref{Resource Estimation}.

\subsection{Mathematical Concept}
\label{Mathematical concept}

The central idea is expressing polynomial functions with binary inputs. For binary digits $x_0,...,x_{n-1}\in\{ 0,1\}$, the number $x_{n-1}...x_0$ has value
\begin{equation}
\label{binary number}
    x_{n-1}...x_0=\sum _{j=0}^{n-1}2^jx_j.
\end{equation}
In some sense, $x_{n-1}...x_0$ represents the linear function $x$ over an exponentially large domain $\{ 0,...,N-1\}$ in terms of a linear number of binary inputs. When taking an exponential of this binary number, we have
\begin{equation}
\label{exponential of binary number}
    \alpha ^{x_{n-1}...x_0}=\prod _{j=0}^{n-1}\alpha ^{2^jx_j}.
\end{equation}
In analogy to a quantum circuit, each $\alpha ^{2^jx_j}$ corresponds to a single qubit rotation gate on qubit $j$ (i.e. multiplication only occurs on the $|1\rangle$ state of qubit $j$).

Meanwhile, the square of this binary number has representation
\begin{equation}
\label{binary number squared}
    \left( x_{n-1}...x_0\right) ^2=\sum _{j=0}^{n-1}4^jx_j+\sum _{j<k}^{n-1}2^{j+k+1}x_jx_k
\end{equation}
since $x_j^2=x_j$. This demonstrates a representation of the quadratic function $x^2$ over an exponentially large domain in terms of a linear number of binary inputs and a quadratic number of binary pairs. Taking an exponential of this expression, we will accrue terms of the form $\alpha ^{2^{j+k+1}x_jx_k}$ which correspond to single qubit controlled rotations (i.e. multiplication only occurs on the $|11\rangle$ state of qubits $j$ and $k$). We can continue this argument to express state-preparation circuits for exponentials of larger $d$ degree polynomials in terms of $O(n^d)$ $(d-1)$ -- qubit controlled rotation gates.

\subsection{Polynomial Phase State-Preparation}
\label{Polynomial phase state-preparation}

\begin{figure}
\input{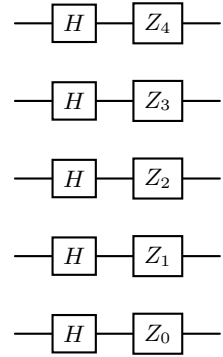}
\caption{State-preparation circuit for the 5-qubit linear phase state $\frac{1}{\sqrt{32}}\sum _{x=0}^{31} e^{i\alpha x}|x\rangle$.}
\label{complex linear phase}
\end{figure}

We now translate the above argument to a tangible circuit which state-prepares
\begin{equation}
\label{complex monomial phase state}
    |\psi\rangle =\frac{1}{\sqrt{N}}\sum _{x=0}^{N-1}e^{i\alpha x^d}|x\rangle.
\end{equation}
For $d=1$, using the $Z$-rotations from Equation \ref{Z_m} we can describe the phase gradient tensor product as
\begin{equation}
\label{diagonal phase gradient}
    Z_{n-1}\otimes ...\otimes Z_0=\sum _{x=0}^{N-1}e^{i\alpha x}|x\rangle\langle x|.
\end{equation}
Applying this to a Hadamard state as in Figure \ref{complex linear phase} gives a standard linear phase state.

\begin{figure*}
\input{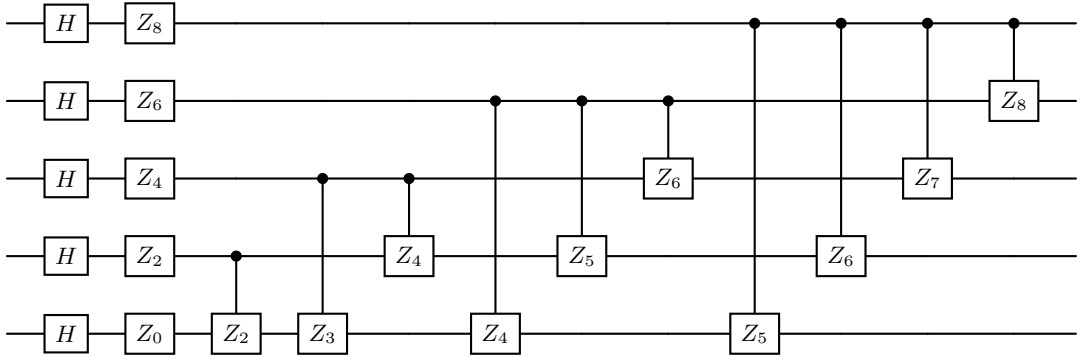}
\caption{State-preparation circuit for the 5-qubit quadratic phase state $\frac{1}{\sqrt{32}}\sum _{x=0}^{31} e^{i\alpha x^2}|x\rangle$.}
\label{complex quadratic phase}
\end{figure*}

\begin{figure}
\input{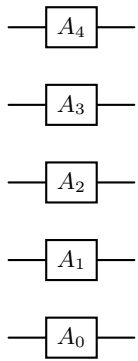}
\caption{State-preparation circuit for the 5-qubit linear phase state $\frac{1}{\sqrt{32}}\sum _{x=0}^{31} \alpha ^{x}|x\rangle$.}
\label{linear real phase}
\end{figure}

By direct analogy with Equation \ref{binary number squared}, we can encode a quadratic phase state by placing controlled rotations according to the equation. This can be further extended to higher order polynomial phases as in Figure \ref{complex quadratic phase}. Since each of the rotations occur on the $z$-axis of the Bloch sphere, these polynomial phases may be applied to any initial state (not just the uniform superposition) with no impact to subnormalization.

\subsection{Exponential State-Preparation}
\label{Exponential state-preparation}

Now we shift our attention to state-preparing stretched exponential states, otherwise real analogues of the previous circuits. We first consider a state-preparation scheme for the exponential state
\begin{equation}
\label{real exponential state}
    |\psi\rangle\propto\sum _{x=0}^{N-1}\alpha ^x|x\rangle
\end{equation}
where $|\alpha |<1$. This necessitates creating a gradient state using $Y$-rotations in the form of $A_m$ from Equation \ref{A_m}:
\begin{equation}
\label{A rotations linear state}
    A_{n-1}\otimes ...\otimes A_0|0...0\rangle =\left(\prod _{k=1}^n\frac{1}{\sqrt{1+\alpha ^{2k}}}\right)\sum _{x=0}^{N-1}\alpha ^{x}|x\rangle.
\end{equation}

\subsection{Gaussian State-Preparation}
\label{Gaussian state-preparation}

\begin{figure*}
\input{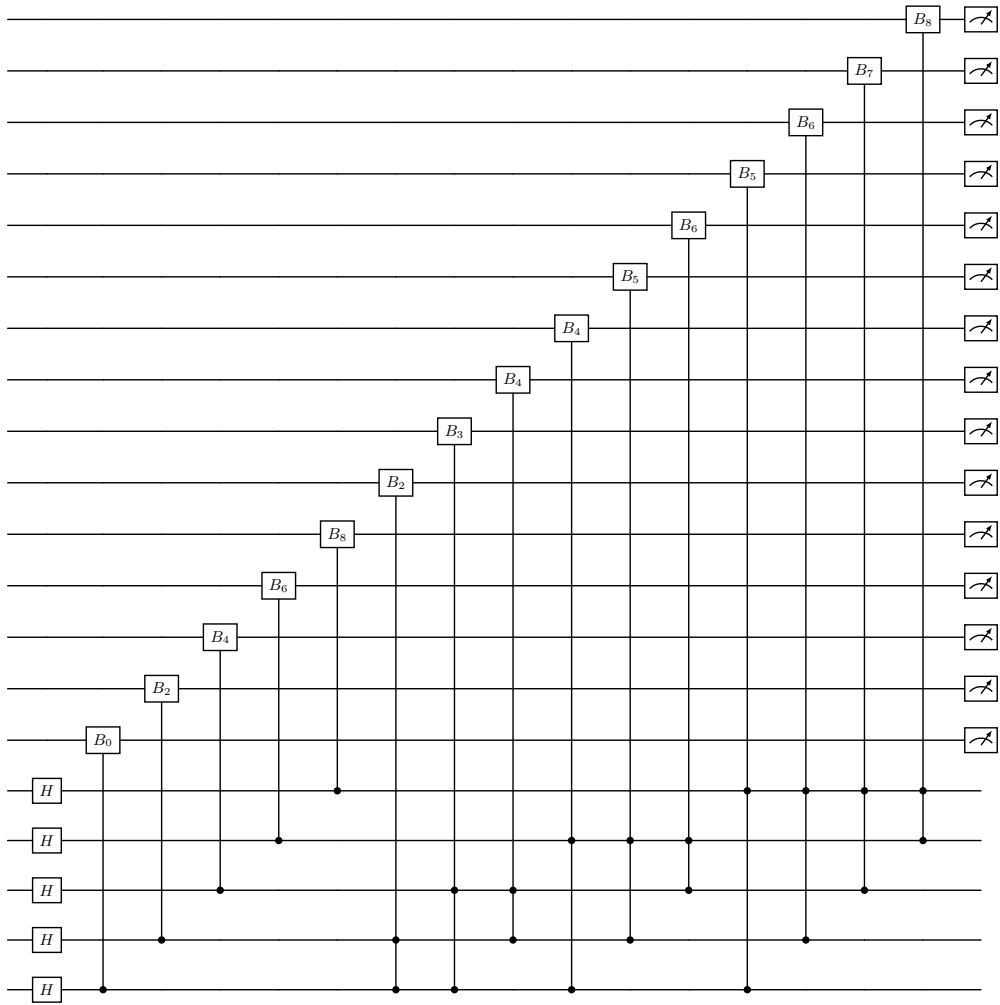}
\caption{State-preparation circuit for a 5-qubit half-Gaussian state $\sum _{x=0}^{31}\alpha ^{x^2}|x\rangle$. All qubits begin in the $|0\rangle$ state and the measured qubits are considered successful if they end in the $|0\rangle$ state. The bottom five qubits are the data qubits.}
\label{unoptimized half gaussian}
\end{figure*}

The extension of the previous exponential state-preparation to a Gaussian state-preparation scheme is not as trivial as preparing higher order complex polynomial phases was, precisely because the $Y$-rotation is not a diagonal operator. Instead, we will use ancilla qubits to \emph{block-encode} \cite{clader22} a real analogue of the diagonal operators which appear in the previous section. Using the $B_m$ matrices from Equation \ref{B_m}, let $CB_m$ represent a controlled $Y$-rotation with target on the top qubit and control on the bottom qubit with the matrix representation
\begin{equation}
\label{controlled B rotation}
    CB_m=\begin{pmatrix} 1 & 0 & 0 & 0 \\ 0 & \alpha ^{2^m} & 0 & -\sqrt{1-\alpha ^{2^{m+1}}} \\ 0 & 0 & 1 & 0 \\ 0 & \sqrt{1-\alpha ^{2^{m+1}}} & 0 & \alpha ^{2^m}\end{pmatrix}.
\end{equation}
Now, this controlled rotation block-encodes a real diagonal matrix with a ratio of $\alpha ^{2^m}$ between bottom-right and top-left entries of the block. A similar doubly-controlled $B_m$ matrix with controls on qubits $j$ and $k$ and target on an ancilla block-encodes a $4\times 4$ diagonal matrix where the bottom-right entry has a ratio of $\alpha ^{2^m}$ to the other terms. To tensor and multiply these matrices together as we did in the previous section, we must use a separate ancilla for each gate as in Figure \ref{unoptimized half gaussian}. This results in a block-encoding of a diagonal matrix $D=\sum _{x=0}^{N-1}\alpha ^{x^2}|x\rangle\langle x|$, which allows us to apply a \emph{half Gaussian-window} to any initial state. In the case of the half-Gaussian, we simply apply this to a uniform superposition Hadamard state.

To convert this half-Gaussian circuit to a full symmetric Gaussian state-preparation, per the previous discussion we construct a circuit to prepare an $n$-qubit state written as
\begin{equation}
\label{final target state}
    |\psi\rangle\propto \sum _{x=0}^{N-1}\alpha ^{(x-(N-1)/2)^2}|x\rangle.
\end{equation}
To create this state, we apply the half-Gaussian window to an initial exponential state $\sum _{x=0}^{N/2-1}\alpha ^x|x\rangle$ created purely by $A_m$-type $y$-rotations (this initial exponential state accomodates the linear term in the exponent of equation \ref{final target state}, effectively shifting the inputs by 1/2). This gives us an $n-1$-qubit state proportional to $\sum _{x=0}^{N/2-1}\alpha ^{(x+1/2)^2}|x\rangle$; applying a Hadamard gate to an empty top data qubit gives us two sequential copies of this state. We use $n-1$ CNOT gates with open control on the top qubit and targets on the other $n-1$ data qubits to flip the first half of the state and create a fully symmetric Gaussian from Equation \ref{final target state}.

\begin{figure}
\input{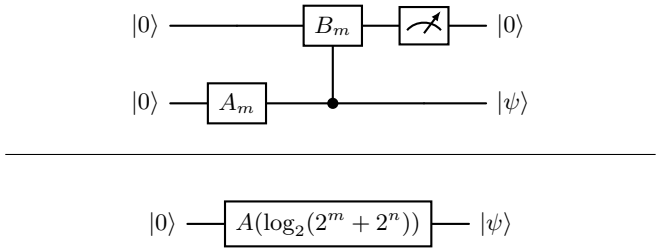}
\caption{Equivalent circuits of a preparation of a one-qubit state $|\psi\rangle \propto\left( |0\rangle +\alpha ^{2^m+2^n}|1\rangle\right)$. This will be used to simplify the circuit in Figure \ref{unoptimized half gaussian}. (\emph{Top}) Beginning in $|00\rangle$, first apply $A_m$ on the second qubit and $CB_n$ with control on the second and target on the first. Then measure the second qubit; success is determined by a $|0\rangle$ measurement. The result is a one-qubit state $|\psi\rangle$. (\emph{Bottom}) Beginning in $|0\rangle$, a single $y$-rotation $A\left(\log_2(2^m+2^n)\right)$ reproduces the same state $|\psi\rangle$.}
\label{block encoded real rotation}
\end{figure}

\begin{figure*}
\input{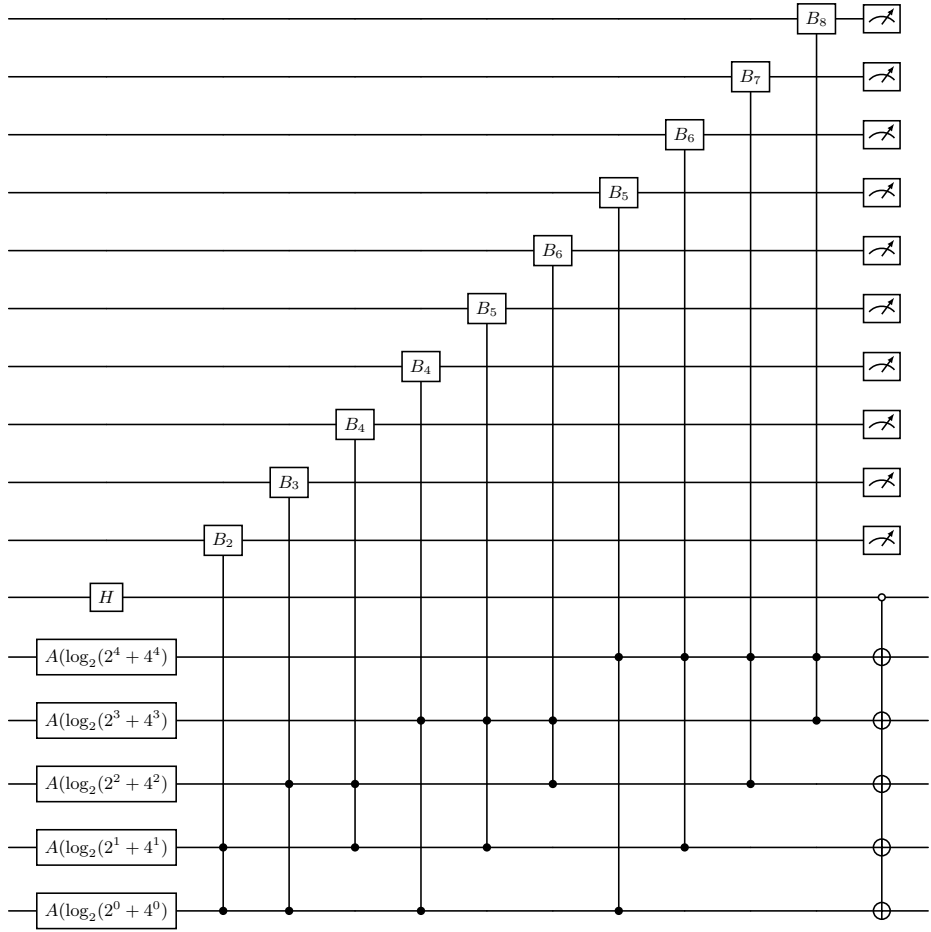}
\caption{State-preparation circuit for a 6-qubit Gaussian state $\sum _{x=0}^{63}\alpha ^{(x-32+1/2)^2}|x\rangle$. Notice that in comparison to Figure \ref{unoptimized half gaussian} we have both shifted the index by $1/2$ and symmetrized the distribution by applying a Hadamard and open control CNOT gates controlled on an extra data qubit.}
\label{full gaussian state}
\end{figure*}

We can further simplify these circuits and boost subnormalization by incorporating the single-qubit controlled rotations into the initial state. Since the initial state created by $B_m$-type rotations applies multiplicative factors of $\alpha ^{2^k}$ to the $|1\rangle$ state of the $k^\text{th}$ qubit and the $A_m$-type controlled rotations apply multiplicative factors of $\alpha ^{4^k}$ to the $|1\rangle$ state of the $k^\text{th}$ qubit, we can simply add exponents to recover this new set of $y$-rotations. Figure \ref{block encoded real rotation} illustrates how a two-qubit circuit combining a $A_m$ gate and a controlled $B_n$ gate applied to $|00\rangle$ with a $|0\rangle$ measurement on the top qubit is equivalent to a single gate $A\left(\log _2(2^m+2^n)\right)$ applied to $|0\rangle$. We display the updated symmetrized Gaussian state-preparation circuit in Figure \ref{full gaussian state}.

While this circuit has the evident downside of requiring $(n-1)(n-2)/2$ persistent ancilla qubits, there are only $(n-1)(n-2)/2$ controlled rotations and the rotation angles are simple to compute. Furthermore, as we will find in the next section, the abundance of persistent ancilla does not meaningfully impact the subnormalization of the state-preparation.

    \section{Circuit Optimizations}
\label{Circuit Optimizations}

We make a variety of observations to greatly simplify the first-pass circuit of Figure \ref{full gaussian state}.

\subsection{Optimizing Error Allocation}
\label{Optimizing error allocation}

First, consider the task of allocating error tolerances to the various rotation gates of the circuit to optimize T-cost. Assuming a finite circuit to approximate an infinite Gaussian, one could in principle parameterize the total state error $\epsilon (\{ \delta _i\} ,\{\delta _{jk}\} )$ by the error allocated to each individual rotation gate and optimize the T-cost of the circuit (proportional to $\sum _{i=1}^n\log\frac{1}{\delta _i}+\sum _{j<k}2\log\frac{1}{\delta _{jk}}$) subject to the constraint $\epsilon (\{ \delta _i\} ,\{\delta _{jk}\} ) =\epsilon$. In practice, these expressions for total error can be difficult to procure for large states and ultimately do not amount to much improvement (we have estimated the benefit to be a $0.5\%$ reduction in T-depth). We anecdotally find success applying uniform errors among both the uncontrolled and two-qubit controlled rotations, with the uncontrolled rotations being twice as accurate as the controlled rotations.

\subsection{Eliminating Angles Below Threshold}
\label{Eliminating angles below threshold}

We now tackle the nuanced question of representing an infinite normal distribution with a finite quantum state. Suppose we wish to prepare an infinite Gaussian with discretization established by $\alpha$ and suppose the circuit is infinite. If we establish an error threshold $\delta$ on each gate, then we need only keep the first $n$ qubits such that the uncontrolled rotation gates are more than $\delta$ from the identity matrix, i.e. if $\alpha ^{2^{n-1}+4^{n-1}}>\delta$ or otherwise \begin{equation}
n=\left\lfloor\log _2\left(\sqrt{1+\frac{4\log\delta}{\log\alpha}}-1\right)\right\rfloor.
\end{equation}

Meanwhile if $\alpha$ is close enough to 1, we may be able to remove controlled rotations dependent on qubits at the bottom of the stack and replace certain uncontrolled rotations with Hadamard matrices. Indeed, we can remove qubit $k$ if its maximal controlled rotation (the rotation controlled on qubit $k$ and $n$) induces a multiplication $\delta$-close to the identity; otherwise if $1-\alpha ^{2^{k+n-1}}<\delta$. Thus we can eliminate controlled rotations on the bottom 
\begin{equation}
    k=\left\lfloor\log_2\frac{\log\delta}{\log\alpha}\right\rfloor +1-n
\end{equation} qubits.

\subsection{Re-using Ancilla Qubits}
\label{Re-using ancilla qubits}

\begin{figure*}
\input{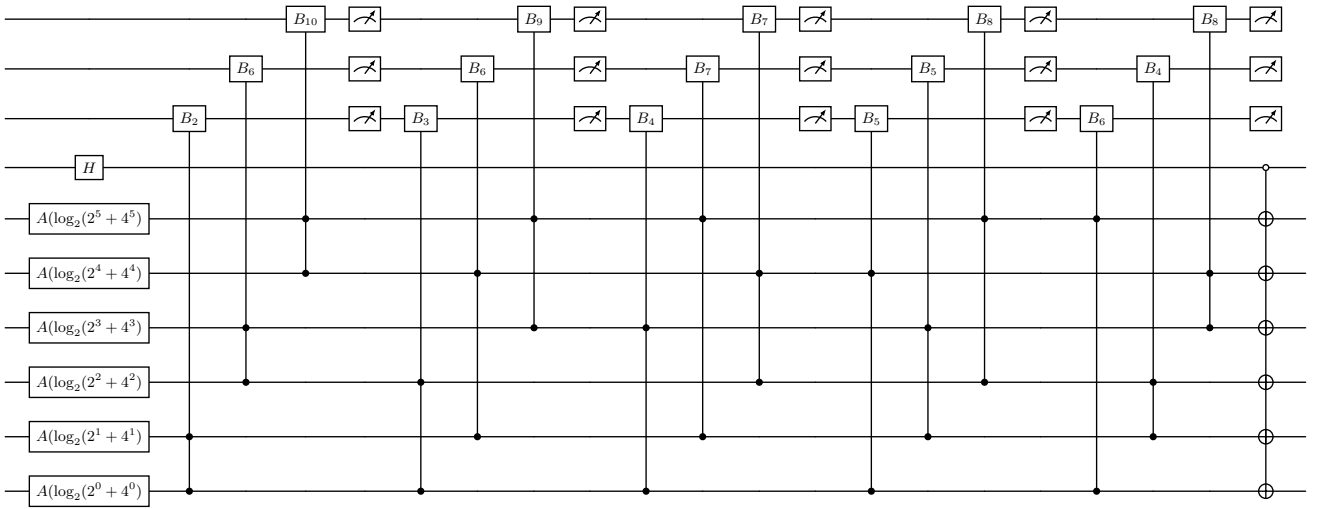}
\caption{State-preparation circuit for a 7-qubit Gaussian state $\sum _{x=0}^{127}\alpha ^{(x-64+1/2)^2}|x\rangle$ re-using ancilla qubits}
\label{optimized circuit}
\end{figure*}

A prohibitive aspect of the circuit in Figure \ref{full gaussian state} is the requirement of $(n-1)(n-2)/2$ persistent ancilla. Fortunately, since each ancilla qubit is separately acted on by a single two-qubit controlled-rotation \emph{we can re-use ancilla qubits after a successful measurement}. For example, we could simply use one ancilla and alternate controlled rotations with measurements $(n-1)(n-2)/2$ times. However, to minimize T-depth we will instead use $(n-1)/2$ ancilla and take measurements between each of the $n-2$ layers ($n-1$ if $n$ is even) of $\lfloor (n-1)/2\rfloor$ controlled rotations. We can then organize rotation gates such that within each layer the $n$ controls precisely occupy each of the $n$ data qubits and do not overlap. This means the controlled rotations in each layer can be executed simultaneously since they all occupy separate qubits, significantly curtailing T-depth. An example of a six-qubit Gaussian state-preparation scheme in Figure \ref{optimized circuit} illustrates one such pattern.

\subsection{Optimizing Measurement Order}
\label{Optimizing measurement order}

With these sequential measurements taking place and ancilla qubits being reused, we no longer need to execute the entire circuit to arrive at a `bad' measurement. Moreover, we can order the layers of rotations between measurements to optimize the total expected number of T-gates. Let $n_0$ be the T-depth of the initial state-preparation and let $n_k$ be the T-depth of each subsequent layer. Meanwhile, let $p_k$ be the probability that the $k^\text{th}$ measurement leads to $|0\cdots0\rangle$ in the ancilla register. If all measurements in the state-prep procedure are successful, the expected T-depth is simply $\sum _{k=0}^n n_k$. Conversely, the expected T-depth of a failed measurement procedure is $n_0+n_1+\frac{1}{1-\prod _{k=1}^np_k}\sum _{k=2}^nn_k\left(\prod _{j=1}^{k-1}p_j-\prod_{j=1}^np_j\right)$. Since the total probability of success is given by $p_s=\prod _{k=1}^np_k$ and the probability of failure is $p_f=1-p_s$, we can compute the total expected T-depth by combining these quantities
\begin{equation}
\label{total expected T-depth}
    \frac{1}{p_s}\left(n_0+\sum _{k=1}^n\left[ n_k\prod _{j=1}^{k-1}p_j\right]\right).
\end{equation}

With this expression, we can re-order rotation layers to optimize for T-depth. Intuitively, we would like to execute layers with larger T-depth and lower probability of success \emph{first} so that we are less likely to encounter a bad measurement at the end of the computation and waste the resources spent to that point. Since all controlled rotations will have the same approximate cost (since they are all being approximated to the same error) we need only focus on the probabilities to optimize the ordering. 

    \section{Resource Estimation}
\label{Resource Estimation}

Armed with all of these optimizations, we are now prepared to estimate the resources required to prepare a Gaussian state. As discussed in Section \ref{Gaussian state-preparation}, there is inherent ambiguity in presenting results for state-preparing an approximation to an infinite distribution like the Gaussian. We will clearly state the parameter space of the following figures.

To calculate expected T-depth for these figures, in addition to varying independent parameters we will vary a \emph{gate error threshold} $\delta$ (different from the resulting error of the state-preparation $\epsilon$). Using this error threshold, we will simulate a rotation gate $A$ with approximate error $\delta$ in a random direction on the Bloch sphere and allocate exactly $1.15\log _2(1/\delta )+9.2$ T-gates (rather than inconveniently simulating each rotation with a different decomposition). We then simulate the circuit in Figure \ref{optimized circuit} with these approximate rotations to recover an approximate Gaussian state which we then calculate the accuracy of by direct $L^2$ norm comparison to an ideal state. To compute T-depth, we create a list of pairs of $n-1$ elements without repetitions and randomly sort it into $n-2$ (or $n-1$) groups of $\lfloor (n-1)/2\rfloor$ elements; these correspond to control qubits of the doubly-controlled rotation gates. Since we approximate each of these rotations to accuracy $2\delta$ per Section \ref{Optimizing error allocation}, they each cost $2.3\log _2(1/\delta )+22.4$ T-gates, and since each rotation in the group acts on a different set of qubits this is also the T-depth of the whole layer. We can then simulate probability of successful measurement for each of these rotation layers and order the layers with increasing probability. We can then use Equation \ref{total expected T-depth} (adding $1.15\log_2(1/\delta )+9.2$ to the first layer for the $A_m$ rotations) to compute the total expected T-depth.

\begin{figure}
% \begin{center}
\includegraphics[scale=0.14]{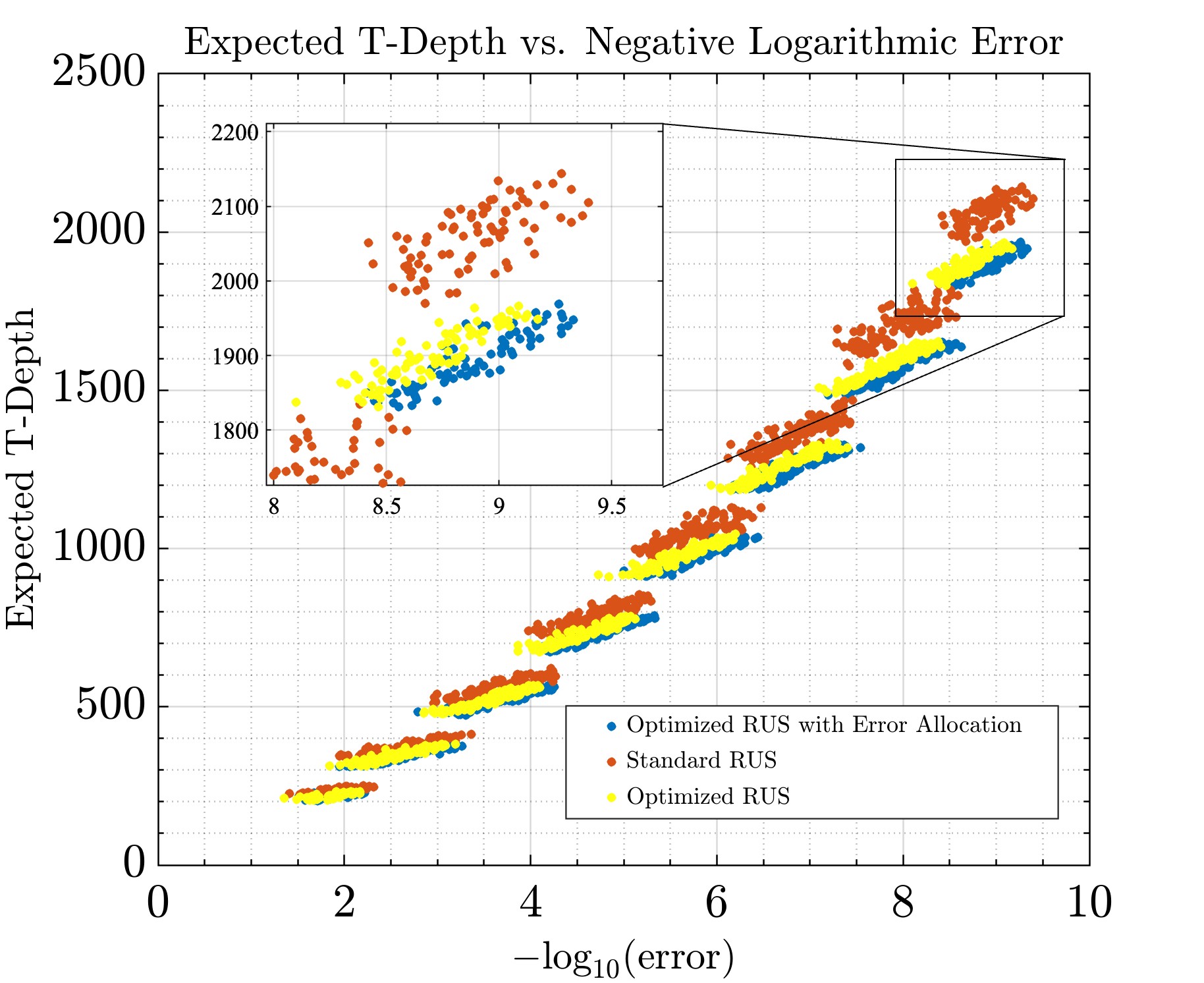}
% \end{center}
\caption{Expected T-depth of Gaussian state-preparation where $\alpha$ varies with the gate error parameter $\delta$ such that $\lVert A_1-XH\rVert =\delta$. Red indicates standard RUS, yellow is optimized RUS, and blue is optimized RUS with improved error allocation.}
\label{matlab plots}
\end{figure}

First, we present a T-depth graph with a single independent variable to help visualize the optimizations we discussed in Section \ref{Circuit Optimizations}. For this graph we choose a gate error threshold $\delta$ as our independent parameter and we choose $\alpha$ such that $\lVert A_1-XH\rVert =\delta$, i.e. that the bottom-most rotation gate is significant enough to be approximated with rotation synthesis such that no gates are eliminated from choosing $\alpha$ too close to 1. Using $\delta$ and this chosen $\alpha$, we compute a threshold number of qubits for the state-preparation and prepare a finite Gaussian over these qubits; the results are plotted in Figure \ref{matlab plots}. For comparison we consider three separate schemes (standard RUS, order-optimized RUS, and order-optimized RUS with optimal error allocation) to illustrate the improvements offered by the optimizations from the previous section. As we can see, error allocation (i.e. doubling the precision of the $A$ rotations compared to the $B$ rotations) offers only a slight improvement, while optimally arranging layer ordering is much more significant (estimated $7\%$ reduction in T-depth over random RUS). Notice the stratification of the graph which occurs as the number of qubits discretely increases.

\begin{figure*}
\begin{center}
\includegraphics[scale=0.255]{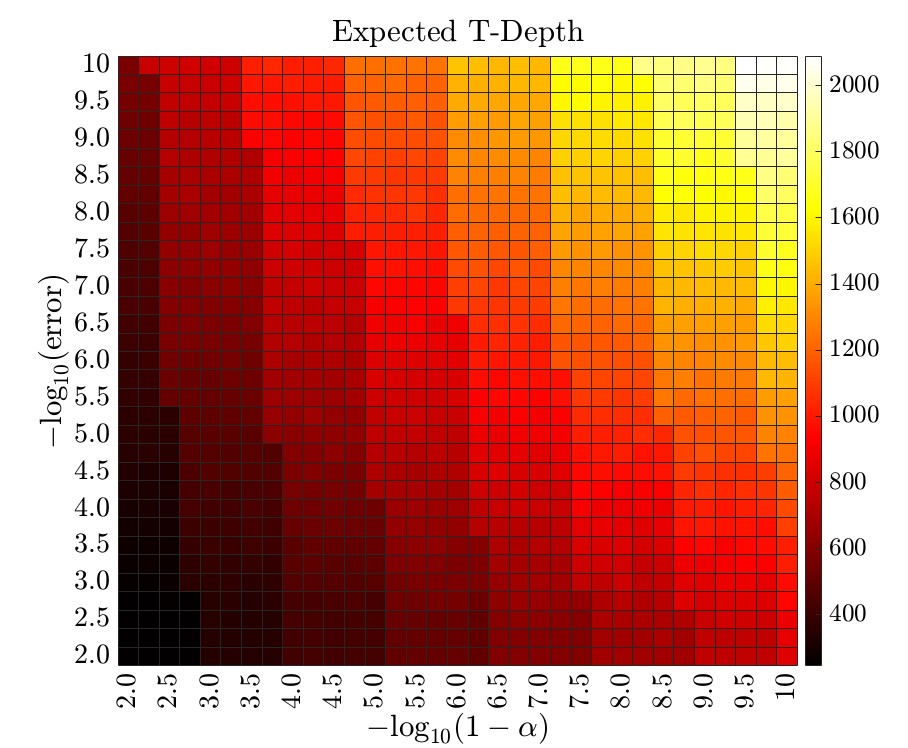}\hspace{1cm}\includegraphics[scale=0.21]{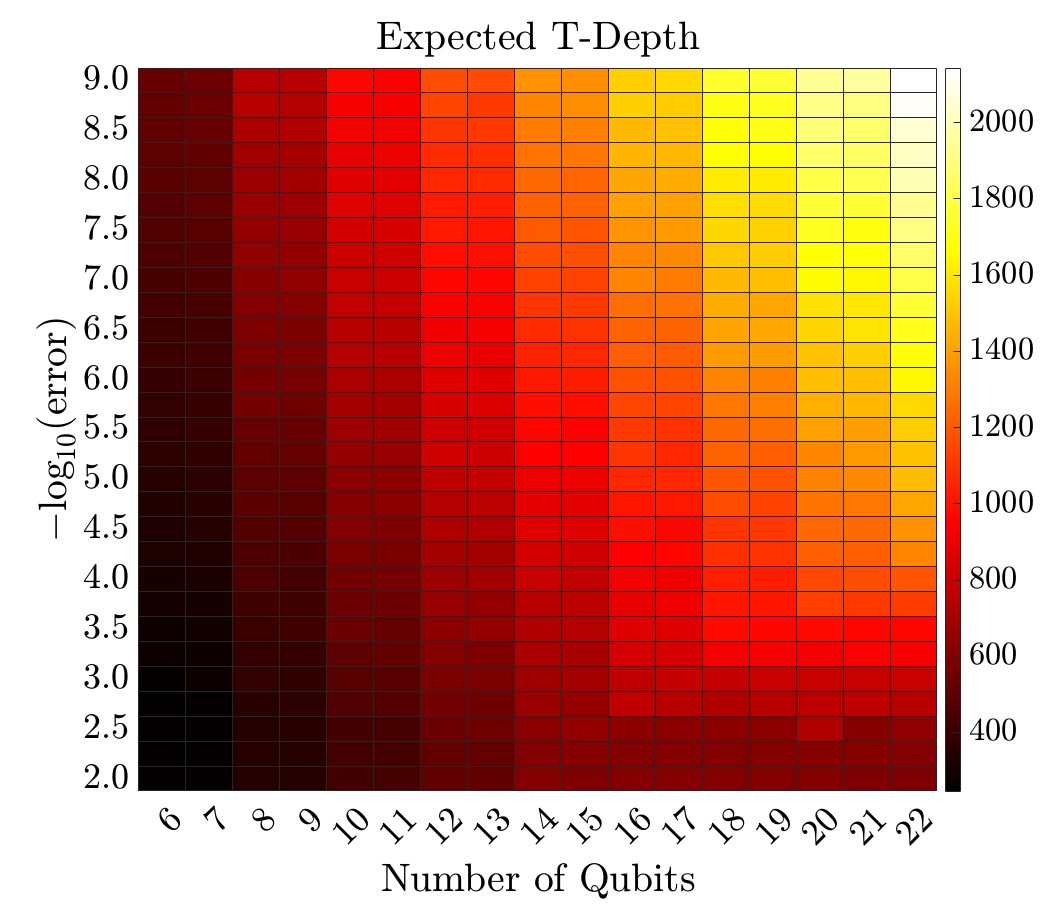}
\end{center}
\caption{Plots of expected T-depth of Gaussian state-preparation circuits for a variety of parameter spaces. (\emph{Left}) Gaussian state-preparation circuit for variable $\alpha$ and total error $\epsilon$. Number of data qubits required for each state-preparation range from $5$ in the bottom left corner (i.e. $\alpha =0.99$, $\epsilon =0.01$) to $19$ in the top right corner (i.e. $\alpha =1-10^{-10}$, $\epsilon =10^{-10}$). (\emph{Right}) Finite Gaussian state-preparation for fixed $\beta\approx 1.3\times 10^{-14}$ with variable number of qubits and total error. The target distribution captures $4\sqrt{2}$ standard deviations of the infinite normal distribution.}
\label{heatmap plots}
\end{figure*}

Next, we observe the expected T-depth of a Gaussian state-preparation with variable total error $\epsilon$ (which is a function of individual gate error $\delta$) and $\alpha$ as depicted in Figure \ref{heatmap plots} (the previous plot in Figure \ref{matlab plots} has a domain that traces out a one-dimensional path in Figure \ref{heatmap plots}). As we can see, there is still a discrete stratification based on the number of qubits required to prepare the Gaussian state; for reference a Gaussian state-preparation with $\alpha =1-10^{-10}$ and $\epsilon =10^{-10}$ requires 19 data qubits to produce. These circuits are fairly efficient at only 2,000 expected T-depth in the most extreme parameter regime.

Finally, we consider a plot comparing the present state-preparation algorithm to the rejection sampling algorithm from Lemieux \textit{et.\ al.} \cite{lemieux24}. This paper documents the Toffoli cost of a rejection sampling algorithm for preparing a finite Gaussian state over a fixed interval, increasing the sampling frequency with increasing qubit count. To compare, we use the $\beta$ formalism from Section \ref{Gaussian state-preparation} such that the target finite Gaussian captures $4\sqrt{2}$ standard deviations of an infinite Gaussian; the results are plotted in Figure \ref{heatmap plots}. In the top corner of the heatmap we see a circuit producing a 22-qubit Gaussian state approximated to $L^2$ error $10^{-9}$ has about 1,900 T-depth. Compare this to Figure 11 in \cite{lemieux24} which documents a Toffoli count of about 50,000 for their rejection sampling algorithm (200,000 T-count). \\

    \section{Conclusion}
\label{Conclusion}

In this paper we have documented a new technique for loading Gaussian distributions onto a quantum computer. This method has several significant advantages over other Gaussian state-preparation methods in the literature, including its aesthetic simplicity, almost nonexistent classical overhead, efficiency in T-depth even for high fidelity states, and low ancilla qubit cost.

\begin{figure*}
\input{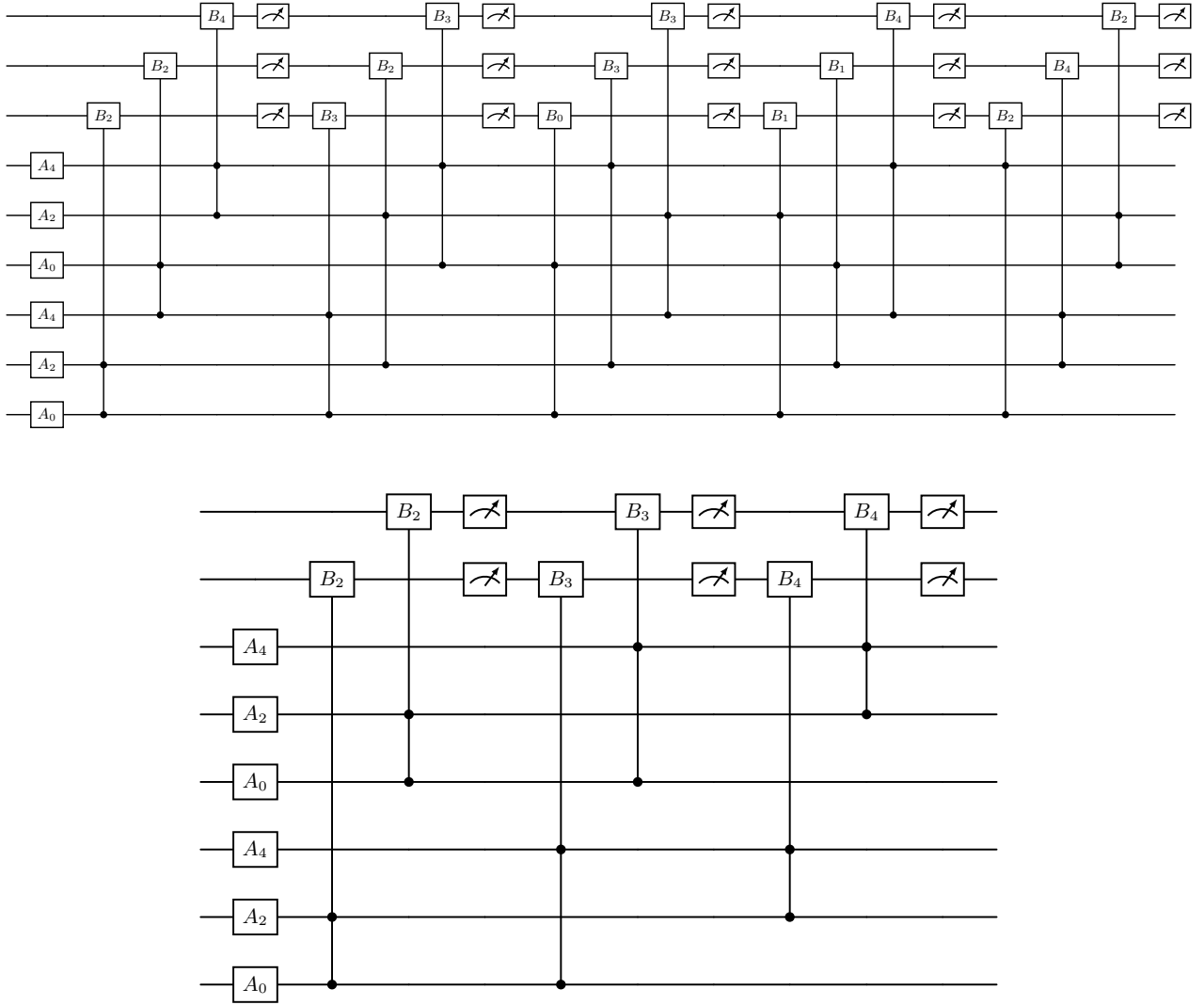}
\caption{State-preparation circuits for a 6-qubit two-dimensional Gaussian distribution over the upper-right quadrant (\emph{Top}) $\sum _{x,y=0}^7\alpha ^{-(x^2+y^2)}|x\rangle |y\rangle$ (\emph{Bottom}) $\sum _{x,y=0}^7\alpha ^{-(x^2+xy+y^2)}|x\rangle |y\rangle$. The top three data qubits correspond to the $|x\rangle$ register and the bottom three data qubits correspond to the $|y\rangle$ register. Notice that the top figure prepares a two-dimensional Gaussian distribution with no covariance and predictably does not have any entangling operations between $|x\rangle$ and $|y\rangle$ registers.}
\label{two dimensional gaussian}
\end{figure*}

We have also commented on its potential to generalize, for instance to preparing quantum states with complex polynomial phase or stretched-exponential states with degree $d>2$. The arguments in Section \ref{State-preparation circuits} conveniently carry over to state-preparation procedures for multi-dimensional Gaussian distributions; a Gaussian with zero covariance between variables can be prepared with a stack of non-interacting circuits similar to Figure \ref{two dimensional gaussian} while a Gaussian state with nontrivial covariance will require controls which span the different input registers. These multi-dimensional Gaussian distributions frequently arise in lattice cryptography problems \cite{chen24}. 

It is the hope of the authors not only that the method described above will eventually replace current popular techniques such as Kitaev-Webb \cite{kitaev08} and other machine learning approaches \cite{manabe24}, but also that the aesthetics of the circuits above may inspire more novel algorithms for quantum state-preparation. The landscape of quantum algorithms is still broadly unexplored and it is simply too early to coalesce subroutines like state-preparation around generic methods when so much progress might be made investigating bespoke techniques for commonly occurring states.

    %============================== 
    
    \section*{Acknowledgments}

    All authors acknowledge support from the Defense Advanced Research Projects Agency under Air Force Contract No. FA8702-15-D-0001. Any opinions, findings and conclusions or recommendations expressed in this material are those of the authors and do not necessarily reflect the views of the Defense Advanced Research Projects Agency.
    
    \bibliography{main}

\end{document}